\documentclass[aps,prd,twocolumn,groupedaddress,nofootinbib]{revtex4-2}

\usepackage{graphicx} % including PostScript
\usepackage[utf8]{inputenc}
\usepackage{color}
\usepackage{hyperref}
\usepackage{amsmath,amssymb}
\usepackage{amsfonts}  
\usepackage{mathrsfs}
\usepackage{subcaption}
\usepackage{pifont}
\usepackage{enumerate} % enumeration
\usepackage{hhline}  %for command \hhline
\usepackage{multirow} %
\usepackage{booktabs} %for commands such as \toprule
\usepackage{tabulary}
\usepackage[compat=1.1.0]{tikz-feynman}
\usepackage{nicefrac}
\usepackage{braket}
\usepackage{comment}
\usepackage{xspace}
\usepackage{cancel}
\usepackage{arydshln}
\usepackage{tikz}
\usetikzlibrary{positioning}
\usepackage[normalem]{ulem}
\usepackage{xcolor}
\usepackage{slashed}

\usepackage{caption}
\usepackage{ragged2e}
\makeatletter
\long\def\@makecaption#1#2{%
  \vskip\abovecaptionskip
  \small
  \sbox\@tempboxa{#1: #2}%
  \ifdim\wd\@tempboxa >\hsize
    \justifying #1: #2\par
  \else
    \global\@minipagefalse
    \hbox to\hsize{\hfil\box\@tempboxa\hfil}%
  \fi
  \vskip\belowcaptionskip}
\makeatother

\newcommand{\D}{{\rm d}}
\newcommand{\E}{{\rm e}}

\newcommand{\LyA}{\textrm{Lyman-$\alpha$ }}
\newcommand{\mLyA}{m_{\textrm{Ly}\alpha} }
\newcommand{\mDM}{m_\chi}
\newcommand{\nEQ}{n^\text{eq}}

\begin{document}

\title{Light Dark Matter from Self-cooling Dark Sectors}
\author{Esau Cervantes, Andrzej Hryczuk, Stefan Lederer}
\affiliation{National Centre for Nuclear Research, Pasteura 7, 02-093 Warsaw, Poland}

%\emailAdd{esau.cervantes@ncbj.gov.pl}
%\emailAdd{andrzej.hryczuk@ncbj.gov.pl}
%\emailAdd{stefan.lederer@ncbj.gov.pl}

\date{\today}

\begin{abstract}
Light thermally produced dark matter is subject to strong bounds stemming from its free-streaming impact on suppressing structure formation. In this paper we show that these limits are significantly alleviated if the dark sector undergoes self-cooling, a new mechanism for lowering the temperature of the dark sector plasma through cannibal self-interactions during freeze-in production. We find that the \LyA bounds can be modified by a few orders of magnitude in the sub-MeV region and frozen-in dark matter saturating the observed energy density can be as light as  $\approx 1.9 (0.7)\text{\,keV}$, compared to the $\approx 5.7 (1.9)\text{\,keV}$ warm dark matter limits determined from simulations. Interestingly, a secondary phase of cooling due to simultaneously efficient inverse cannibal- and decay-processes can dominate the modifications of \LyA bounds, rather than self-thermalization via cannibal-reactions itself.

\end{abstract}
\maketitle

%%%%%%%%%%%%%%%%%%%%%%%%%%%%%%%%%%%%%%%%%%%%%%%%%%%%%
\section{Introduction}
\label{sec:intro}
%%%%%%%%%%%%%%%%%%%%%%%%%%%%%%%%%%%%%%%%%%%%%%%%%%%%%

Cosmological and astrophysical observations established the existence of an additional source of gravitational potential beyond the one coming from baryonic matter. This is attributed to dark matter (DM), most properties of which are yet to be understood. In particular, the mass of objects constituting dark matter can span many orders of magnitude from  $10^{-21}$\,eV to $10^{37}$\,kg (see \cite{Cirelli:2024ssz} for a recent review). This range becomes smaller if the DM has particle nature and the mechanism establishing its present day abundance is thermal, $\mathcal{O}(1)$\,keV to $\mathcal{O}(100)$\,TeV for standard cosmology. The upper bound stems from unitarity limit on the annihilation cross section~\cite{Griest:1989wd}, while the lower limit comes from formation of large-scale structures in the Universe, in particular via the observations of the \LyA absorption lines produced by intergalactic neutral hydrogen in the spectra of distant quasars~\cite{McQuinn:2015icp}. In contrast to the cosmic microwave background measurements, which constrain the DM energy density, $\Omega h^2=0.1198\pm 0.0012$ \cite{Planck:2018vyg}, the \LyA observations lead to upper bounds on the DM velocity distribution.

The existing measurements interpreted within the warm dark matter (WDM) model, i.e. an energetic weakly interacting massive particle (WIMP) undergoing freeze-out to saturate the DM abundance, are respected for $m_{\rm WDM}\gtrsim 5.7\text{\,keV}$~\cite{Irsic:2023equ} or as low as $1.9\,\text{keV}$ assuming colder, later reionization histories with reduced IGM thermal smoothing, which leaves more room for warm-dark-matter-induced suppression~\cite{Garzilli:2019qki}. 

Multiple studies have recast \LyA limits beyond the thermal WDM template, either by matching suitably defined free-streaming or velocity measures, or by computing the linear matter-power suppression from the non-thermal phase-space distribution. This has been done for broad classes of non-cold dark matter spectra~\cite{Heeck:2017xbu,Murgia:2018now,Ballesteros:2020adh,Dienes:2021cxp}, for FIMP freeze-in through decays and scatterings~\cite{Kamada:2019kpe,Decant:2021mhj,DEramo:2020gpr,DEramo:2025jsb,Fuyuto:2024oii}, 
cannibal freeze-out~\cite{Garny:2018byk}, resonant sterile-neutrino production~\cite{Bringmann:2022aim,Kasai:2025xaw}, Shi-Fuller type production~\cite{Vogel:2025aut} and for mixed freeze-in/superWIMP production~\cite{Decant:2021mhj,Zhao:2026wxi}. More general likelihood-based treatments likewise show that \LyA data can be applied to feebly interacting and mixed cold-plus-non-cold scenarios, provided the shape of the small-scale power suppression is treated consistently~\cite{Hooper:2022byl}. The connection between these cases is that the production history fixes a momentum scale, typically set by the temperature of the SM bath at production, which then controls the late-time velocity dispersion of the dark matter population.

It follows that for DM freeze-in proceeding from a dark sector (DS) bath with a temperature $T_\text{DS}<T$, the limits are weaker and lower DM masses remain in agreement with large-scale structure formation observations. However, any prior thermal contact between the visible and dark sectors means that $T_\text{DS}$ and $T$ are related and turn out to be of similar order. For substantial deviation from simple red-shifting, the dark sector needs to incorporate either a huge number of relativistic degrees of freedom or another process which lowers $T_\text{DS}$ after SM and DS have thermally decoupled.  

Here we consider a scenario where DM is produced from a decay within a DS involving additionally a dark mediator, an unstable real scalar $\phi$ or vector boson $A_\mu$. Absent a stabilizing symmetry, there is no reason for such real mediators not to undergo number-changing scattering processes such as $2\to 3$ as well. Indeed, cubic operators are a common feature of a multitude of models (see e.g.~\cite{Batell:2010bp,Hochberg:2014kqa,Bernal:2015bla,Bernal:2015lbl,Tulin:2017ara}), making cannibalization a generic feature of the mediator.

Without the cannibalization process, the DM production becomes a sequential freeze-in scenario \cite{Hambye:2019dwd}, $ {\rm{SM}} \rightarrow \phi,A \rightarrow \text{DM} $.
Number-changing self-interactions such as $2\to 3$ processes allow the transfer of excess heat into mediator number density, an effect which we refer to as \textit{self-cooling}. Previous studies \cite{Bernal:2020gzm,Cervantes:2024ipg} suggest that substantially lowered $T_\text{DS} \ll T$ are possible in such models.

The goal of this work is to demonstrate and quantify the impact of a self-cooling DS on lower DM mass bounds from \LyA observations. We construct analytic approximations of naively expected and maximal limiting cases 
and compare them to more realistic numerical solutions of coupled second-moment Boltzmann equations  (cBE). While freeze-in of DM or DS is being studied intensively, $2\leftrightarrow 3$ processes are regarded far less frequently, despite the fact that cannibal reactions can be sourced by something as generic as a 3-point interaction in the DS. With this work, we highlight that higher-order reactions which affect otherwise conserved thermodynamic quantities (here DS abundance) must not be neglected on generic grounds of QFT power counting arguments without prior consideration of their cosmological relevance.

This paper is structured as follows. We begin by introducing the production and cooling mechanisms in detail in section~\ref{sec:cooling}, including approximate analytic results as far as possible. For concreteness of our numerical analyses, we introduce minimal models for scalar and vector dark mediators in section~\ref{sec:models}. We computed numerical solutions to the set of coupled number and temperature Boltzmann equations of the mediator and discuss our results, including a Monte-Carlo parameter search as well as two-dimensional parameter scans, in section~\ref{sec:results}. We summarize our results and draw conclusions in section~\ref{sec:conclusions}.

%%%%%%%%%%%%%%%%%%%%%%%%%%%%%%%%%%%%%%%%%%%%%%%%%%%%%
\section{Self-cooling of a Dark Sector}
\label{sec:cooling}
%%%%%%%%%%%%%%%%%%%%%%%%%%%%%%%%%%%%%%%%%%%%%%%%%%%%%

Let us assume that the dark sector consists of the DM candidate ($\chi$) and a ``cannibal" field ($\phi$) with both having negligible populations right after the end of reheating. The mediator $\phi$ plays a central role in the dynamics of the mechanism, while $\chi$ is taken to be simply a product of $\phi$ decay without any other impact on DS dynamics and having no direct interactions with the SM sector.\footnote{
    In principle, the reproductive part of the mechanism would work perfectly well without separating the roles of the mediator and DM, making $\phi$ to be both. However, the resulting dark matter would undergo strong self-scatterings, such that its sub-MeV realizations would make such a scenario excluded by observations of DM self-interactions \cite{Randall:2008ppe,Bernal:2025osg}.
}

\begin{figure}[t]
    \centering
    \includegraphics[width=0.98\linewidth]{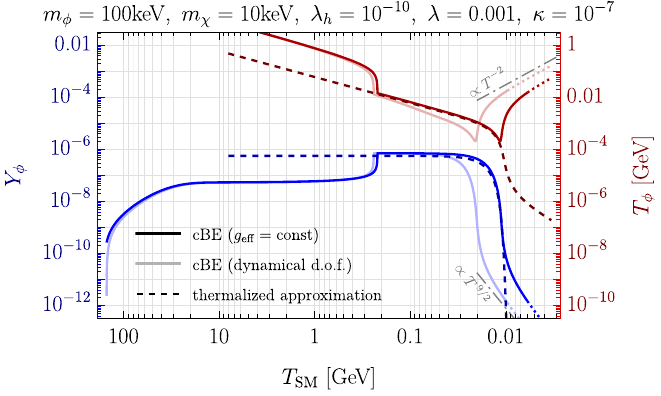}
    \caption{Illustration of the self-cooling mechanism together with comparison of different semi-analytical approximations.
    Abundance $Y_\phi$ (blue) and temperature $T_\phi$ (red) evolution of the dark sector field $\phi$ are plotted over the Standard Model temperature $T_{\rm{SM}}=T$. Numerical cBE solutions are shown with and without $T$-dependent effective degrees of freedom \cite{Drees:2015exa} (lighter and darker solid curves). Thermal $\phi$ evolution obtained from non-relativistic Higgs-decay production are superimposed (dashed). cBE solutions continue beyond their respective kinetic decoupling as dotted curves. 
    }
    \label{fig:EvolutionOverview}
\end{figure}

Fig.~\ref{fig:EvolutionOverview} illustrates the self-cooling mechanism in $\phi$ abundance $Y_\phi$ (blue) and temperature $T_\phi$ (red). The various depicted approximations will be discussed in the remainder of this chapter. 
For now, the focus is the features visible in $Y_\phi$ and $T_\phi$ evolutions. The distinct step-feature in abundance at $T\approx 0.2$\,GeV arises from inverse-cannibal reactions starting to be efficient and once $\phi$ reaches internal thermalization its abundance growth is halted abruptly. $T_\phi$ drops correspondingly but also shows a second, spiked feature at later times caused by the additional interplay of decay processes. Overall, the dark mediator is cooled by roughly an order of magnitude relative to the SM during the thermalization cooling and conserves that temperature ratio until decay sets in.

%----------------------------------------------------
\subsection{Coupled Boltzmann equations}
\label{sec:cooling-BME}
%----------------------------------------------------

In principle, a determination of \LyA bounds on $\chi$ requires a solution of the Boltzmann equations at phase-space level (e.g. by utilizing numerical codes like DRAKE~\cite{Binder:2021bmg} or BEST~\cite{Yoon:2026rce}) for the complete dark sector, i.e. both $\phi$ and $\chi$ simultaneously. However, if the central ingredient of the considered scenario, efficient cannibal reactions $2\phi \leftrightarrow 3\phi$, arises from renormalizable interactions, then by necessity also elastic $2\phi\to 2\phi$ scatterings will be efficient as well. Therefore, the momentum distribution of the efficiently reproducing particle species is continuously kept close to a thermal one. However, at later times both, $2\to2$ and $2\leftrightarrow3$, scatterings can become inefficient, which necessitates a separate consideration, as discussed in Sec.~\ref{sec:KD}. 

Assuming a sufficiently dilute DS, the phase space distribution is well described by a  Maxwell-Boltzmann (MB) distribution, $f_\phi(p,T)\approx n_\phi(T_\phi)/\nEQ_\phi(T_\phi)\times \E^{-E_p/T_\phi}$, where $E_p$ denotes the energy of $\phi$ given momentum $p$ and $T_\phi$  the temperature of the $\phi$ population. 
This considerably simplifies the calculation of DS self-cooling: instead of solving for the full phase-space distribution, it is sufficient to evolve the zeroth and first energy moments of the Boltzmann equation~\cite{Bringmann:2006mu,Binder:2017rgn}. The DM phase-space distribution at a time parametrized by SM temperature $T$ may then be computed from the number density $N_\phi$ and mediator temperature $T_\phi$. Introducing
\begin{align}
    x\equiv \frac{m_\phi}{T},
    ~~
    x_\phi\equiv \frac{m_\phi}{T_\phi}
    .
\end{align}

and using that entropy conservation gives $\D x/\D t=xH/\tilde g(T)$ where $\tilde g = 1+ T/(3g_{\star S})\partial g_{\star S}/\partial T$ and $g_{\star S}$ is the entropy effective degrees of freedom. The coupled Boltzmann equations (cBE) can be written for $Y_\phi\equiv n_\phi/s$ and $T_\phi$
\begin{align}
    \frac{\D Y_\phi}{\D x}
    &=
    \frac{\tilde g}{xHs}C_0[\phi],
    \label{eq:cBE}
    \\
    \frac{\D T_\phi}{\D x}
    &=
    \frac{
        \frac{\tilde g}{xH}C_E[\phi]
        -\frac{3\tilde g}{x}P_\phi
        -\frac{\partial\rho_\phi}{\partial Y_\phi}
        \frac{\D Y_\phi}{\D x}
    }{
        \frac{\partial\rho_\phi}{\partial T_\phi}
    } ,
    \label{eq:cBE_Tphi}
\end{align}
where $C_0[\phi]$ and $C_E[\phi]$ are the zeroth and first energy moments of the collision operator, {\it c.f.}~Sec.~\ref{sec:models}.  The pressure term appears as $P_\phi$, rather than $\rho_\phi+P_\phi$, because $\rho_\phi=\rho_\phi(Y_\phi,T_\phi,x)$ has an explicit $x$-dependence through $s(x)$. The corresponding $\partial\rho_\phi/\partial x$ contribution cancels the term proportional to $\rho_\phi$ in the redshifting part of the energy equation. 
Under the MB ansatz, all moments of the phase-space distribution are fixed by $n_\phi$ and $T_\phi$, such that the zeroth and energy moment equations form a closed system. Hence in the dilute limit, the thermodynamic quantities entering Eq.~(\ref{eq:cBE},~\ref{eq:cBE_Tphi}) are
 \begin{align}
    \rho_\phi
    &=
    n_\phi
    \left[
        3T_\phi+
        m_\phi \frac{K_1(x_\phi)}{K_2(x_\phi)}
    \right],
    \\
    P_\phi
    &=
    n_\phi \, T_\phi\,,
\end{align}
where $n_\phi^{\rm eq}(T_\phi)=\frac{m_\phi^2T_\phi}{2\pi^2} K_2(m_\phi/T_\phi)$ is the equilibrium number density.

Numerical solutions to these cBE are shown as solid curves in Fig.~\ref{fig:EvolutionOverview} and constitute our most accurate results. Darker colours assume constant effective degrees of freedom, $\tilde g=1$, in the same cBE setup and are thus expected to agree well with our analytic approximation derived below.

\bigskip

To build some expectation, we construct analytic solutions for the late time behaviour of the cBE solutions. To do so, we take $\tilde g=1$.
In the relativistic limit, $x_\phi\ll 1$, the Bessel functions may be expanded around small arguments, 
\begin{align}\label{eq:GammaEff}
       \Gamma_{\rm eff} = \,-\frac{C_0^{\rm dec}[\phi]}{n_\phi}
    =\,
    \Gamma_\phi
    \frac{K_1(x_\phi)}{K_2(x_\phi)}
    \approx\,
    \Gamma_\phi\frac{m_\phi}{2T_\phi}
    ,
\end{align}
whilst also $\rho_\phi \approx 3n_\phi T_\phi$, $P_\phi \approx n_\phi T_\phi$ and
\begin{align}
    C_E^{\rm dec}[\phi]
    &=
    -\Gamma_\phi m_\phi n_\phi .
\end{align}

Applied to Eq.~\eqref{eq:cBE_Tphi}, expanding to leading order in $T_\phi$ and assuming additionally $H=H_\star/x^2$ and $\D x/\D t = x H /\tilde g \approx x H$ yields:
\begin{align}
    \frac{\D T_\phi}{\D x}
    \approx
    -\frac{T_\phi}{x}
    +
    \frac{\Gamma_\phi m_\phi}{3xH}
    ,
\end{align}
which admits the solution
\begin{align}
    T_\phi(x)
    \approx
    \frac{\Gamma_\phi m_\phi}{9H_\star}x^2
    +
    \frac{\text{const}}{x}.
\end{align}
Thus, the late-time attractor during radiation domination behaves as $ T_\phi\propto x^2 $.
The yield equation in the same limit becomes
\begin{align}
    \frac{\D Y_\phi}{\D x}
    \approx
    -\frac{\Gamma_\phi}{xH}
    \frac{m_\phi}{2T_\phi}Y_\phi 
\end{align}
and substituting $T_\phi \approx  \Gamma_\phi m_\phi x^2/(9H_\star)$ gives $Y_\phi\propto x^{-9/2}$. 
In total, find that, as long as kinetic equilibrium is maintained, the late-time mediator evolution will in fact not be an exponential depletion but merely a polynomial one.
To guide the eye, we include both late-time predictions in arbitrary normalization in Fig.~\ref{fig:EvolutionOverview} (dot-dashed gray lines). They correctly predict the qualitative late-time behaviour of the numerical cBE solutions.

%----------------------------------------------------
\subsection{Approximate analytic self-cooling }
\label{sec:analytic}
%----------------------------------------------------

If the mediator bath remains in thermal equilibrium for an extended period of time during the DS self-thermalization (around~$T\approx 0.2$\,GeV in Fig.~\ref{fig:EvolutionOverview}) thermodynamics describes its expected evolution and amount of self-cooling. In particular, since $2\leftrightarrow3$ processes conserve the total energy within the $\phi$-sector, the change in temperature due to chemical equilibration is easily obtained via energy conservation from freeze-in to the time of when thermal equilibrium is reached, 
$\rho^\text{FI}(x_\phi^\text{FI}) = \rho^\text{eq}(x_\phi^\text{eq})$. Here, $\rho^\text{eq}_\phi=\rho_\phi|_{n=n^\text{eq}}$.
Explicitly, this relates the number density and temperature obtained by FI, $n_\phi^\text{FI}$ and $x_\phi^\text{FI}$, to the temperature of the equilibrated mediator $x_\phi^\text{eq}$ at the same point in time:
    \begin{align}
        n_\phi^\text{FI}\times &\left[\frac{3}{x_\phi^\text{FI}} + \frac{K_1(x_\phi^\text{FI})}{K_2(x_\phi^\text{FI})}\right]  
         \nonumber  \\  & =
        \frac{\frak{g}_\phi m_\phi^3 }{2\pi^2 (x_\phi^\text{eq})}  \left[ \frac{3}{x_\phi^\text{eq})}K_2(x_\phi^\text{eq}) +  K_1(x_\phi^\text{eq})\right]  
        .
    \label{eq:chemEqfull}
    \end{align}
    
The ratio $x_\phi^\text{FI}/x_\phi^\text{eq}$ sets the strongest cooling accessible from internal chemical equilibration of $\phi$ alone. 
The mediator abundance is obtained as $n_\phi^\text{eq}(x_\phi^\text{ana})/s$. The DS energy density from non-relativistic Higgs decay  serves as a boundary condition.

Note, that the above considerations rely on the assumption of having sufficient energy in the system to reach chemical equilibrium. 
Formally, the steps of mediator production, thermalization and decay are treated separately in this approach. However, since we neglect both backreactions, $\phi \phi \to h$ as well as $\bar\chi\chi\to\phi$, and cannibal-reactions conserve DS energy density, it is sufficient to have well separated mediator production and decay.
While the abundance evolution in Fig.~\ref{fig:EvolutionOverview} appears almost unremarkable once mediator-decay sets in, $T\sim 0.02$\,GeV, the temperature evolution is more instructive. Both, numeric and analytic solutions, see a second brief phase of cooling, however they separate once $2\to3$ reactions decouple and the numerical evolution falls out of chemical equilibrium. In contrast, chemical equilibrium is baked into Eq.~\eqref{eq:chemEqfull} and the analytic approximation cools much more and well into the non-relativistic regime. We refer to this cooling by reproduction of decaying modes via efficient $2\to3$ scatterings, 
\begin{equation}
\Gamma_{2\to 3} = \frac{C_0^{(2\to3)}}{n_\phi } = \left<\sigma v\right>_{2\to 3}n_\phi \gtrsim \Gamma_\text{dec}\gg H,
\end{equation} 
as \emph{decay cooling}.   
As the $\phi$ population continuously decays to $\chi$, any losses are immediately regenerated from the thermal bath to the thermal equilibrium value. Since this rapidly, in fact exponentially, cools the DS, also the abundance depletes. 
It constitutes a significant or even dominant contribution to the total DS cooling in large parts of parameter space. 

Note that a decay cooling would not have been observed, had one incorrectly assumed an instantaneous decay approximation:
For $\Gamma_\text{dec}\gg\Gamma_{2\to3}\gtrsim H$, the DS cannot maintain chemical equilibrium and $Y_\phi$ decreases, decoupling $\Gamma_{2\to3}$ despite the fact that the reproduction initially remains an efficient process.\footnote{
    In rapid sequential freeze-in, the number changing rate is governed by $\Gamma_\text{FI}/\Gamma_\text{dec}$ as any produced $\phi$ decays ``immediately" (on cosmological scales). Here, it would be conceivable to have a hierarchy $\Gamma_\text{FI}, \Gamma_{2\to 3}\gg \Gamma_\text{dec}\gg H$, in which case still chemical equilibrium is dynamically maintained despite rapid production and depletion of $\phi$ itself.
    }

Within the relativistic regime, the solution of Eq.~\eqref{eq:chemEqfull} simplifies to 
${T_\phi}^\text{eq} \approx \sqrt{\pi}\sqrt[4]{n_\phi^\text{FI} T_\phi^\text{FI} / \frak{g}_\phi}$. 
The initially frozen-in DS abundance is set by the Higgs portal coupling square, $n_\phi^\text{FI}\propto \Gamma_\text{FI}\propto g_h^2$, yet its temperature is set by the energy injected per particle and, thus, depends primarily on $E_h/m_\phi \approx m_h/m_\phi$. Since in the relativistic regime also $\nEQ(x_\phi^\text{eq})\sim (x_\phi^\text{eq})^{-3}\propto (n_\phi^\text{FI})^{3/4}$, we find that $x_\phi^\text{FI}/x_\phi^\text{eq}\approx \sqrt{g_h}$ and for well separated reproduction and decay processes, the maximal mediator abundance scales as $\max \{Y_\phi\}\propto g_h^{3/2}$. 
Non-relativistic Higgs decay indeed dominates the production mechanism, hence the dashed curves in Fig.~\ref{fig:EvolutionOverview} agree well with the thermalized region, $T\sim 0.2-0.01$\,GeV, as long as $\tilde g=1$ is also implemented in the numerical solutions (darker solid curves). 

Fig.~\ref{fig:EvolutionOverview} also includes numeric cBE solutions with temperature-dependent effective degrees of freedom (lighter solid curves). $\tilde g\neq1$ causes the mediator to decay at roughly a factor 2 to higher temperatures. Because the QCD phase transition ($T\sim 0.1$\,GeV) occurs systematically within the critical time window for GeV--keV DS masses, it is not possible to obtain reliable estimates at a better than $\mathcal{O}(1)$ precision from Eq.~\eqref{eq:chemEqfull}.

Maintaining efficient number-changing interactions well into the non-relativistic regime leads to internal freeze-out of the mediator \cite{Carlson:1992fn} via $3\to2$ reactions. This would heat the DS, thus reverting the previously achieved cooling, and is contrary to the intended purpose of the mechanism discussed in this work. We remark that there is a minor parameter space slice where the mediator undergoes $3\to2$ freeze-out, yet the decay cooling still outweighs the heating from cannibal freeze-out but do not investigate this scenario further.

Before moving on, we comment on the requirements for self-cooling to take place.\footnote{If the initially frozen-in mediator abundance is too large, chemical equilibration will heat rather than cool the DS.} There are only three conditions to the chemical equilibration of $\phi$: first of all, self-reproduction must become efficient, $\Gamma_{2\phi\to3\phi}\gtrsim H$, which demands sufficiently large self-couplings for a given frozen-in number density. Once efficient, the self-reproduction sources a rapid, exponential growth of $n_\phi$. Secondly, there must be sufficient kinetic energy in the dark sector to feed reproduction all the way to complete thermalization. In cases where there is insufficient kinetic energy to produce $\phi$, the reproduction will cease and will not support full internal thermalization. Lastly, the lifetime of $\phi$ must be sufficiently long for the equilibration to complete. In many models, this last condition is easily satisfied by choosing sufficiently small decay rates. However, the decay rate may not always be governed by an independent parameter, as is the case in the SU($N$) example introduced in section~\ref{sec:model-SUN}.

%----------------------------------------------------
\subsection{\LyA bounds}
\label{sec:LyA}
%----------------------------------------------------

Our overall goal is to quantify the impact of DS self-cooling on DM \LyA bounds. 
The DM temperature, or equivalently its mean squared effective velocity $\left< p^2 /\mDM^2\right>$, is a convenient proxy to estimate the impact of DM on structure formation, circumventing the need for expensive N-body simulations of each particle physics model. 
We employ velocity dispersion matching \cite{Bae:2017dpt,Kamada:2019kpe} to translate experimental bounds from WDM simulation results to our models. Since back-reactions are negligible for dilute DM, $\partial_a f_\chi = C[f_\phi,f_\chi] \approx C[f_\phi]$, its phase-space Boltzmann equations can already be integrated in time ($x$ or scale factor $a$) and moment-expanded prior to numerical computations. The squared mean velocity is given by
    \begin{align}
        \frac{1}{m_\chi^2}\left< p^2 \right> &= 
        \frac{
	           \int_{x_i}^x\D r\,\frac{ n_\phi(r)}{\bar{H}(r) } \frac{r^4}{x_\phi(r)} \left[ 3 + \left(\frac{m_\phi^2}{4m_\chi^2}-1\right) \frac{K_3(x_\phi)}{K_2(x_\phi)}\right] 
            }{
	           x^2 ~\int_{x_i}^x\D r\,\frac{n_\phi(r)}{\bar{H}(r) } r^2 \frac{K_1(x_\phi(r))}{K_2(x_\phi(r))}
            },
    \label{eq:vDM}
    \end{align}
and its number density at every point in time $x$ by 
    \begin{align}
        n_\chi(x)=2\times \frac{\Gamma_{\phi\to\chi\chi}}{2x^3} \int_{x_0}^x\D r\,\frac{n_\phi(r)}{\bar H(r)}r^2\frac{K_1(x_\phi(r))}{K_2(x_\phi(r))}
        .
        \label{eq:nDM}
    \end{align}

The velocity dispersion matching is a simple approximation scheme for the structure formation impact of DM but performs well for unimodal DM momentum distributions, in particular if they do not differ significantly from thermal ones~\cite{Dienes:2021cxp}. We confirmed by numerical computation of $f_\chi[n_\phi,x_\phi]$ that our DM momentum distribution satisfies this condition and hence we expect velocity dispersion matching to perform well, with merely percent-level deviations from more elaborate techniques \cite{Decant:2021mhj,Dienes:2021cxp}. Unimodal distributions are generally expected for DM produced via decay from a bath in kinetic equilibrium but become increasingly flattened (``box-shaped") if mother particles are relativistic. \LyA bounds on the mean momentum can then be expressed as a lower bound on $m_\chi$ depending on $\sqrt{\left<p^2\right>}$,
    \begin{equation}
        m_\chi \geq  7.45\cdot10^6\,\times\,\sqrt{\left<p^2\right>}\,\times\,\left(\frac{m_\text{WDM}^{\text{Ly-}\alpha}}{1\,\text{keV}}\right)^{4/3}
    ,
    \label{eq:LyA}
    \end{equation}
where the lower bounds on WDM are taken from existing simulations. We primarily use $m_\text{WDM}^{\text{Ly-}\alpha}=5.7$\,keV~\cite{Irsic:2023equ}, but also show results for a more conservative bound 1.9\,keV \cite{Garzilli:2019qki}.

As an analytic approximation of the naively expected impact of self-thermalization alone on the parameter space, we again take the relativistic approximation for $H$ and $n_\phi$ and define $x_\phi/x \equiv \xi$ for late times where DM is produced. This again takes $\tilde g =1$ and assumes unperturbed redshifting in the DS for relativistic temperatures. Expanding the integrands in Eq.~\eqref{eq:vDM} and \eqref{eq:nDM} to leading orders in $x_\phi$ yields  $\left.\left<v^2\right>\right|_x=\frac1x \frac{9}{\xi^3}\left(\frac{m_\phi^2}{4m_\chi^2}-1\right)$. 

Some care must be taken when defining the expected corrections to the \LyA bounds due to number-changing self-interactions because we intend to compare two valid DM models, both of which are to saturate the energy density bound. Therefore, the appropriate comparison of a given model point of  $g_h$ undergoing self-thermalization must be to a model of larger $\bar g_h \equiv g_h^{3/4}$ and identical $\mDM$ but disabled $2\leftrightarrow3$ scatterings.

%----------------------------------------------------
\subsection{Treatment after kinetic decoupling}
\label{sec:KD}
%----------------------------------------------------

As the mediator decays, its abundance drops rapidly until eventually also elastic $2\to2$ scatterings decouple and the cBE (\ref{eq:cBE},~\ref{eq:cBE_Tphi}) as well as Eqn.~(\ref{eq:nDM},~\ref{eq:LyA}) are no longer applicable. We assume instantaneous kinetic decoupling at time $x_\text{kd}$ set by $\left<\Gamma_{2\to2}\right>|_{x_\text{kd}}=H(x_\text{kd})$. In Fig.~\ref{fig:EvolutionOverview}, the numeric cBE solutions are shown even for $x>x_\text{kd}$ as dotted lines for illustrative purposes, despite being unreliable here.
It is not necessary to obtain a best-possible estimate of $x_\text{kd}$ as modifications obtained after kinetic decoupling give rise to $\mathcal{O}(10\%)$ corrections in the mean effective velocity, $\sqrt{\left<p^2\right>/m_\chi^2}$. However, it remains crucial to include a reasonable ($\sim\mathcal{O}(1)$ accurate) endpoint $x_\text{kd}$. The reason is that under the assumption of eternal kinetic equilibrium the decaying mediator bath will heat rapidly and freeze-in ever hotter DM modes. Such spurious hot modes will lead to a growth of the mean effective velocity with the cBE evolution endpoint $x_{f,\text{cBE}}$ for $x_{f,\text{cBE}}\gg x_\text{kd}$.
In the scalar model, $\Gamma_{2\to2}$ may be simplified by the trivial quartic contact interaction to a good approximation. We terminate the numerical solution of cBE at $x_{f,\text{cBE}}\equiv x_\text{kd}$ and compute the remaining contribution under the assumption of a free-streaming unstable mediator,
\begin{align}
    f^\text{fs}_\phi(a_\text{kd},p) &= \frac{n_\phi(a_\text{kd})}{n_\phi^\text{eq}(a_\text{kd})} \,\E^{-x_\phi(a_\text{kd})\sqrt{1+p^2/m_\phi^2}}
    ,
\\
    \partial_a  f^\text{fs}_\phi(a,p) &=-\frac{m_\phi}{E_p}\frac{\Gamma_{\phi}}{a\,H} \, f^\text{fs}_\phi(a,p)
    ,
\\
    f_\phi^\text{fs}(a_f,p) &= f_\phi^\text{fs}(a_\text{kd},\frac{a_f p }{ a_\text{kd}})\,\E^{-\int_{a_\text{kd}}^{a_f}\frac{\D a}{a H}\frac{\Gamma_\phi m_\phi}{\sqrt{m_\phi^2+p a_f^2 /a^{2}}}} 
    .
\end{align}

It is efficient to construct a numerical interpolation grid for $f_\phi^\text{fs}$ in $a$ and $p/m$ to avoid stacking numerical integrations in the later computation of $f_\chi$.

Since $f_\phi$ is now a numerical object, it is no longer possible to collapse the calculations of $n_\chi$ and $\left<p^2\right>$ to single-integral expressions. Instead, there is technically a total of four stacked integrations remaining:
\begin{align}
    n\, {\delta}\!\left[ p^j \right]\!|_{x_f} =& \int_0^\infty \frac{\D p\,p^{2+j}}{2\pi^2}\int^{a_f}_{a_\text{kd}} \frac{ \D a}{a H} \frac{1}{2E_\chi(q)}
    \notag\\&\times
    \int\! \D\Pi^{\phi,\bar\chi}_\text{LIPS} \, \overline{|\mathcal{M}_{\phi\to\bar\chi\chi}|^2} f^\text{fs}_\phi(a,p_\phi)
    .
\end{align}

By $n\,{\delta}\! \left[ p^j \right]$, we refer to the contribution obtained solely from evolution beyond kinetic equilibrium and $E_\chi(q)$ indicates the energy at intermediate time $a$ with $q=p\,a_f /a$ and the equation is derived for dilute DM, $f_\chi\ll 1$. The Lorentz-invariant phase space includes the four--dimensional Dirac-delta, which cancels out some integrals, but $f_\phi$ involves another integration. In kinetic equilibrium, the next step is to evaluate $\int\D\Pi^{\phi,\bar\chi}_\text{LIPS} = \int_{E_-}^{E_+}\D E_\phi$ with $E_\pm=E_\pm(a,p)$. Instead, we proceed by rewriting the DM momentum integral into comoving momenta to make it independent of $a$ and the proceed first with the integration of $p_{\bar\chi}$ and $q_\chi$. Only the relativistic energy denominator and the flux factor $1/(2E_\chi p_\chi)$ remain after resolving the Dirac delta, and we are left with an integral
\begin{align}
    \label{eq:ndeltaFS}
    n\, {\delta}\!\left[ p^j \right]\!|_x =& \int^{a_f}_{a_\text{kd}} \frac{\D a}{a H} \frac{1}{2E_\chi(q)}
    \notag\\&\times
    \int_0^\infty \!\D p_\phi\, F(p_\phi) \int_{q_-}^{q_+}\!\D q\, \frac{q^{1+j}}{E_\chi(q,a)}
    ,
\end{align}
where $F(p_\phi)$ for our case is
\begin{equation}
    F(p_\phi)\equiv \frac{\D p}{\D q}\left(\frac pq\right)^{2+j}\frac{\int\!\D\Omega~\overline{|\mathcal{M}|^2}}{32\pi^4}=\frac{\kappa^2 m_\phi^2 \epsilon^2}{8\pi^3}\left(\frac{a_f}{a}\right)^{3+j}
    .
\end{equation}
and the integration boundaries are given by $q_\pm = \frac12 \left( p_\phi \pm \epsilon E_\phi\right)$ for 
\begin{equation}
    \label{eq:epsilon}
    \epsilon\equiv\sqrt{1-\frac{4m_\chi^2}{m_\phi^2}} 
    .
\end{equation}

The general integral has a closed-form solution, 
\begin{equation}\label{eq:ComovIntegral}
    \int\!\frac{\D \tilde q ~ \tilde q^{1+j}}{\sqrt{m^2+\tilde q^2}}=\frac{q^{2+j}}{(2+j)m_\chi}
    \,{}_2F_1\!\left( \frac 12 ,\frac{2+j}{2} ; \frac{4+j}{2}; \frac{- {\tilde{q}^2}}{m^2}\right)
    ,
\end{equation}
where the hypergeometric function collapses for $j/2\in \mathbb{N}_0$. 
For $j=0,2$, the final integral in Eq.~\eqref{eq:ndeltaFS} yields 
\begin{align}
    j=0:&~\epsilon p_\phi 
    ,
    \\
    j=2:&~\frac{\epsilon^3  p_\phi }{4}\left [m_\phi^2+p_\phi^2\left(\frac{1}{\epsilon^2}+\frac13 \right)\right]
    .
\end{align}

The number density decomposes simply as a sum into the cBE beyond kinetic equilibrium contributions, $n_\chi(x_f)=\frac{a_\text{kd}^3}{a_f^3} n_\chi(x_\text{kd})+{\delta}\!\left[p^0\right]\!|_{x_f}$, however the second moment involves normalization by the number density,
\begin{equation}
    \left<p^2\right>|_{x_f} = \frac{1}{ n_\chi(x_f)} \left[  \frac{a_\text{kd}^5}{a_f^5} n_\chi(x_\text{kd})\left<p^2\right>|_{x_\text{kd}} + n\,{\delta}\!\left[p^2\right]\!|_{x_f}\right]
    \!.
\end{equation}

%%%%%%%%%%%%%%%%%%%%%%%%%%%%%%%%%%%%%%%%%%%%%%%%%%%%%
\section{Dark sector models}
\label{sec:models}
%%%%%%%%%%%%%%%%%%%%%%%%%%%%%%%%%%%%%%%%%%%%%%%%%%%%%

The self-cooling mechanism and thermodynamic considerations introduced above may arise in various models. We present here two different setups where the self-interacting dark mediator is either a real scalar or a non-Abelian gauge boson. The DM candidate is kept as a Dirac fermion in both cases, for simplicity. For the second model, the gauge symmetry is assumed to be broken spontaneously at some higher scale, giving rise to the mediator mass. Further, we do not aim to connect either model to other known experimental anomalies in the SM in this work. 
The non-Abelian model is of interest as sizable cubic mediator interactions are a necessary component and non-Abelian gauge extensions are studied extensively in modern model-building efforts~\cite{Boddy:2014yra,Hambye:2008bq,Arkani-Hamed:2008hhe,Asadi:2021pwo,Biondini:2023vss,Beneke:2024nxh,Abe:2026bci,Ghosh:2026mda}.

%----------------------------------------------------
\subsection{Scalar mediator}
\label{sec:scalarmodel}
%----------------------------------------------------

The possibly simplest realization of the self-cooling scenario uses $\phi$, a real scalar with the portal interaction to the SM Higgs whose cubic self-interaction term was generated as a consequence of a spontaneously broken $\mathbb{Z}_2$ symmetry \cite{Hufnagel:2022aiz}. If coupled to a $\mathbb{Z}_2$ symmetric Dirac fermion $\chi$ the Lagrangian of the model reads:
\begin{align}
    \label{eq:Lscalar}
    \mathcal{L}_s =& \,\mathcal{L}_\text{SM}+ \frac12 \phi(-\Box-m_\phi^2)\phi + \overline{\chi}\left(i\slashed\partial-m_\chi\right)\chi \nonumber \\
     &+ \frac{g_{h}}{2} |H|^2 \phi^2
     + \frac{\sqrt{3\lambda}}{3!}m_\phi\phi^3+\frac{\lambda}{4!}\phi^4
     + \kappa \phi \overline{\chi}\chi
     .
\end{align}

We briefly summarize the main ingredients of the model below and refer to previous in-depth studies of the same model for more details~\cite{Cervantes:2024ipg,Bernal:2025osg}. Self-interactions of the scalar field are allowed via cubic and quartic self-interactions of $\phi$, which are related to one another motivated from spontaneous breaking of a $\mathbb{Z}_2$ symmetry \cite{Cervantes:2024ipg}.\footnote{This relation can of course be relaxed by considering a generic cubic operator $\lambda_3 \phi^3$, without qualitatively affecting the presented mechanism. However, a practical benefit of choosing $\lambda_3\equiv \sqrt{3\lambda}$ is not only reduction of the parameter space of the model, but more importantly that it  allows to factorize the coupling itself from the calculation of the inelastic self-scattering matrix element, $\overline{|\mathcal{M}_{2\to3}|^2}\propto \lambda^3$, greatly reducing computational cost in the numerical scan. Furthermore, we neglect the possibility of a cubic term in the Higgs portal, $\sim |H|^2\phi$, as well as Yukawa to SM fermions. Including the linear coupling to the Higgs boson does not conceptually alter the mechanism. Additional diagrams to the self-interaction amplitude are suppressed by the Higgs mixing while opened decay channels to the SM are suppressed by light fermion Yukawa couplings for sub-GeV mediators.}

The model has 5 free parameters: the two masses $m_\phi$ and $m_\chi$ and three couplings, $g_h$, $\lambda$ and $\kappa$, governing respectively: freeze-in production of $\phi$ from the Higgs, the $2\phi\leftrightarrow3\phi$ cannibal reactions and the decay rate of $\phi$ to DM. The masses are in principle unconstrained (even $m_\phi > m_H$ would be accessible since $HH\to \phi\phi$ upscatterings are included \cite{Cervantes:2024ipg}), however we will focus on a region with $m_h>2m_\phi$ and $m_\phi>2m_\chi\sim\mathcal{O}$(keV-MeV). The former relation ensures sufficient production of $\phi$'s from the Higgs decay without the need of a substantially large portal coupling, $g_h$. The latter allows for the decay of $\phi$ to DM and the choice of $m_\chi$ being light is motivated by observational relevance of the self-cooling mechanism.\footnote{
    There are models in the literature where \LyA bounds are applicable for heavier DM candidates as well, for which an additional phase of self-cooling might be consequential, see e.g.~\cite{Ballesteros:2020adh,Hambye:2020lvy}.
}

The collision moments entering the cBE~(\ref{eq:cBE},~\ref{eq:cBE_Tphi}) can be decomposed as
\begin{align}
    C_0[\phi]
    &=
    C_0^{h\to\phi\phi}
    + C_0^{3\phi\leftrightarrow2\phi}
    + C_0^{\phi\to\bar\chi\chi},
    \label{eq:C0_scalar_decomp}
    \\
    C_E[\phi]
    &=
    C_E^{h\to\phi\phi}
    + C_E^{\phi\to\bar\chi\chi},
    \label{eq:CE_scalar_decomp}
\end{align}
where \(C_E^{3\phi\leftrightarrow2\phi}=0\), since the
\(3\phi\leftrightarrow2\phi\) reactions conserve the total energy
inside the mediator bath. Additionally, the zeroth moment of the cannibal reactions is $C_0^{3\phi\leftrightarrow2\phi} = n_\phi \braket{C^{3\leftrightarrow2}}$,  with $\braket{C^{3\leftrightarrow2}}$ given in App.~A of~\cite{Cervantes:2024ipg}.

Neglecting inverse processes, the Higgs-decay source gives
\begin{align}
    C_E^{h\to\phi\phi}
    &=
    \Gamma_{h\to\phi\phi}\,m_h\,
    n_h^\text{eq}\!\left(\frac{m_h}{T}\right)
    \\&=
    \frac{g_h^2 v_h^2}{32\pi}
    \sqrt{1-\frac{4m_\phi^2}{m_h^2}}\,
    n_h^\text{eq}\!\left(\frac{m_h}{T}\right)
    \label{eq:CE_FI_scalar}
    ,
    \\
    C_0^{h\to\phi\phi}
    &=
    C_E^{h\to\phi\phi}\,
    \frac{2}{m_h}
    \frac{K_1\!\left(\frac{m_h}{T}\right)}
         {K_2\!\left(\frac{m_h}{T}\right)}
    \label{eq:C0_FI_scalar}
    .
\end{align}
Similarly, the mediator decay contribution is 
\begin{align}
    C_E^{\phi\to\bar\chi\chi}
    &=
    -\Gamma_{\phi\to\bar\chi\chi} \, m_\phi \,n_\phi\!\left(x_\phi\right)
  \\&=
    -\frac{\kappa^2 m_\phi^2 \epsilon^3}{8\pi}\,
    n_\phi
    \label{eq:CE_dec_scalar}
    ,
    \\
    C_0^{\phi\to\bar\chi\chi}
    &=
    C_E^{\phi\to\bar\chi\chi}\times
    \frac{K_1(x_\phi)}{m_\phi K_2(x_\phi)}
    \label{eq:C0_dec_scalar}
    ,
\end{align}
with the phase space factor of Eq.~\eqref{eq:epsilon} $\epsilon\approx 1$ in most points of the parameter space. We study degenerate models of $m_\phi=2.01m_\chi$ ($\epsilon=0.10$) separately in section~\ref{sec:results-2d}. Outside of this narrow region, the self-cooling dynamics is almost unaffected by $m_\chi$. Indeed, as long as all processes occur in the relativistic regime $x,x_\phi\ll1$, also $m_\phi$ becomes effectively arbitrary and may always be absorbed into a combination of $g_h$, $\lambda$ and $\kappa$. Since $\chi$ does not take part in reproduction and there is a degeneracy of $\kappa$ and $\epsilon$ in $\Gamma_{\phi\to\chi}$, the DM mass only affects the overall scale of the contribution to the relic abundance and the velocity dispersion, impacting \LyA observations. 
Therefore, out of 5 free parameters of the model, one may focus only on the couplings. 

$g_{h}$ governs the overall energy density in the dark sector, while $\kappa$ mainly determines the time of DM production, which we require to be late enough to allow for a preceding phase of mediator self-interactions.
To avoid thermalization among different sectors, both $g_{h}$ and $\kappa$ must remain small, for keV-scale DM typically $\lesssim \mathcal{O}(10^{-6})$. Finally, $\lambda$ is determining the strength of mediator self-interactions and thus reproduction and self-cooling.

%----------------------------------------------------
\subsection{SU($N$) embedding}
\label{sec:model-SUN}
%----------------------------------------------------

The cubic (and quartic) self-interactions which enable self-cooling are a mandatory ingredient of SU($N$) gauge theories.  This suggests that generic DS models with non-Abelian gauge may exhibit a self-cooling mechanism. For concreteness, assume an $N=2$ symmetry\footnote{We limit ourselves to SU($2$), since the gauge symmetry only impacts numerical $\mathcal{O}(1)$ constants in our calculation.} and, in order to facilitate easy comparison of differences with respect to the previous model, we aim to keep the production mechanism as close to the scalar model as possible. This requires the decay of gauge bosons $A$ to DM, so we assume the dark SU($N$) symmetry to be broken well above the electroweak scale. Additionally, this breaking must introduce an effective vertex to the Higgs boson, allowing for freeze-in of $A_\mu$ from Higgs decays as in the previous section.\footnote{
     One elegant implementation would be a dark-Higgs doublet $\Phi$ which fully breaks the dark SU($2$), which we take to be diagonal in $A^a$, and the UV Higgs-potential $\supset|H|^2|\Phi|^2$ gives rise to the Higgs-portal in Eq.~\eqref{eq:Lsun}. This mechanism would only couple to the longitudinal modes of $A_\mu$ but as all our calculations are unpolarized, we may absorb this at leading order into a redefinition of $g_h$ by 1/3.
} 
The effective Lagrangian for the non-Abelian model takes the form 
\begin{align}
\label{eq:Lsun}
    \mathcal{L}_{\text{SU(}N\text{)}}=&\,\mathcal{L}_\text{SM} 
    -\frac14  F^a_{\mu\nu}F^{a,\mu\nu} -\frac12 (m_A^2-g_h v^2) A^{a}_\mu A^{a,\mu} \nonumber \\
    &+ \frac{g_{h}}{2} |H|^2 A^a_\mu A^{a,\mu} 
    + \overline{\chi}\left(\mathrm{i}\slashed D-m_\chi\right)\chi
    ,
\end{align}
where $F$ is the field strength tensor of $A$. The dark gauge coupling $g$ takes the role of the self-interaction $g\,{\sim}\,\sqrt{\lambda}$ in the scalar model but simultaneously governs the decay to $\chi$, $g\,{\sim}\,\kappa$. It is, however, not mandatory to use $\chi$ as a dark multiplet, since we have broken the gauge symmetry already. 
Instead, one can easily consider another model realization  with an effective vertex of the form $\kappa \overline{\chi}\chi \sum_a A^a_{||}$, where $A_{||}$ denotes the Goldstone-equivalent longitudinal polarization of the massive vector bosons.
In this realization, $g_h$, $g$ and $\kappa$ are once more independent\footnote{
    There are consistency relations among $g_h$, $g$ and $\kappa$, which depend on the symmetry breaking mechanism. If $\kappa$ arises from a heavy particle loop, $\kappa \ll g$. For a dark-Higgs mechanism, $g_h$ is not constrained, since the longitudinal polarization of $A_\mu$ couples directly to $H$. Yet, the origin of $g_h$ from an effective vertex may provide a good reason for why it would be quite small.
} 
and $\Gamma_\text{dec}^{A_{||}\to\bar\chi\chi}=\Gamma_\text{dec}|_\phi$.

In section~\ref{sec:results-SUN} we discuss both the case of SU($2$) charged DM doublets, $\kappa=g$ as well as realizations of independent $\kappa$ and $g$.
Even for the former, production and decay rates in the cBE~\eqref{eq:cBE} are almost identical to the scalar model,
\begin{align}
    \left.\Gamma_\text{FI}\right|_\text{SU($2$)} &= \frac{\frak{g}_A^2}{\frak{g}_\phi^2}\,\left.\Gamma_\text{FI}\right|_\phi
    \label{eq:GammaFIsun}
    \,,
    \\
    \left.\Gamma_\text{dec}\right|_\text{SU($2$)} &=~ T_F \left(\frac{1}{\epsilon^2}-\frac13\right)  \times\frac{g^2}{\kappa^2}\left.\Gamma_\text{dec}\right|_{\phi}
    \label{eq:GammaDECsun}
    \,,
\end{align}
naturally inserting $m_\phi=m_A$ and $x_\phi=x_A$ in all cases and $\frak{g}_\phi=1$, $\frak{g}_A=(2s+1)(N^2-1)=9$. $T_F=\frac12$ in the last line arises from the gauge structure trace for a doublet $\chi$. The inverse cannibal collision term may be obtained analogously to \cite{Cervantes:2024ipg}, where the cross-section in Eq.~(A.15) has to be replaced by its appropriate SU($N$) counterpart, including $m_\phi\to m_A$, $\lambda\to g^2$ and averages of initial spins (which is automated in calcHEP) and colors. 
We performed the color summation decoherently, giving another factor $1/9$ for $N=2$. Note that this assumes non-zero mass-splitting among $A^a$, as otherwise the initial gauge boson two-particle states would decompose into $\text{Ad}\otimes\text{Ad}=1\oplus3\oplus5$ of SU($2$). We take $\delta m_A \ll m_A$, such that decay occurs always before $T_A\sim \delta m_A$. 
The freeze-in production rate changes trivially to account for ${\frak g}_A \neq \frak{g}_\phi$, since Eq.~\eqref{eq:Lsun} couples $H$ to all nine terms $\sum_{a,\mu}(A^a_\mu)^2$.

The crucial difference between the non-Abelian and the scalar model arises from the cubic gauge-boson self-interaction which is now momentum dependent. We obtained $(\sigma v)_{2A\to 3A}$ numerically using calcHEP \cite{Belyaev:2012qa} (version 3.9.2) and performed the thermal average ourselves.
Relative to the scalar model, the momentum dependence enhances the resulting reproduction cross-section $(\sigma v)_{2A\to 3A}$ at high temperatures by $\sim T^{3.5}$.

The strong momentum, and thus temperature, dependence warrants a brief discussion. Its origin is the breaking of SU($N$) itself which spoils cancellations among massless $A$ scattering diagrams. The effective Lagrangian in \eqref{eq:Lsun} includes on the vector boson mass term but ignores any higher-dimension effective operators sourced by the UV-complete theory. Hence, the theory is only valid far below the mass scale of any additional UV-completion ingredients and the strong temperature dependence does not hold infinitely. For sufficiently small gauge couplings, the $2\to2$ and $2\to 3$ scatterings remain perturbative until the UV-completion scale. However, assuming correctness of \eqref{eq:Lsun} for any $g$, the obtained temperature dependence raises concerns of unitarity violation at high temperatures.  The solid-angle integrated cross-sections are not subject to a finite upper bound by unitarity, yet a complete partial-wave analysis of the $2\to N$ self-scattering of $A$ and its unitarization is well beyond the scope of this work. We note that, should the elastic $2\to2$ process (which is leading in orders of $g$) saturate at the unitarity bound at high temperatures, any other scattering processes will necessarily be suppressed \cite{Flores:2024sfy}, including the here crucial $2\to3$ inverse-cannibal process. Hence, a careful analysis could yield strong phenomenological effects.\footnote{
    Elastic $2\to2$ scatterings are subject to inherent complications such as t-channel forward enhancements, though IR divergences are regulated by $m_A$ here.
}

We briefly turn to some generic phenomenology of the non-Abelian model. The enhanced temperature dependence causes a much more sensitive dependence of the time of thermalization on the value of $g$ and often results in immediate thermalization.  For gauge-induced mediator decay, $\kappa\sim g$, too large $g$ may cause $A$ to also decay immediately and the parameter space becomes even more limited, especially for large $m_A$. Above $g>10^{-3}$ \cite{Heeba:2018wtf}, also the free-streaming of keV-scale DM should be considered with some care.

%%%%%%%%%%%%%%%%%%%%%%%%%%%%%%%%%%%%%%%%%%%%%%%%%%%%%
\section{Results}
\label{sec:results}
%%%%%%%%%%%%%%%%%%%%%%%%%%%%%%%%%%%%%%%%%%%%%%%%%%%%%

We solve the cBE~(\ref{eq:cBE},~\ref{eq:cBE_Tphi}) numerically for both models of Sec.~\ref{sec:models}. For the scalar model, to identify interesting regions, we have employed scans in the full 5 dimensional parameter space using Monte Carlo Markov chains (MCMCs) and following that studied systematically variations in 2 numerically most relevant dimensions. For the vector model we restrict ourselves to a single benchmark point as a proof-of-concept of our self-cooling mechanism.

%----------------------------------------------------
\subsection{Dark sector evolution and decay cooling}
\label{sec:results-evolution}
%----------------------------------------------------

\begin{figure*}[t]
    \centering
    \hfill\noindent
    \includegraphics[width=0.49\linewidth]{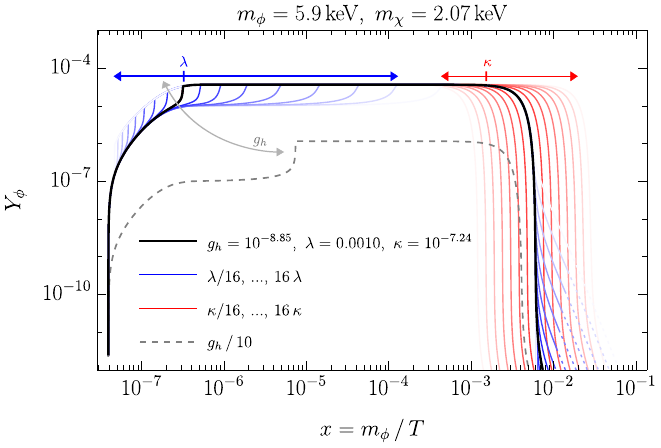}\hfill
    \includegraphics[width=0.49\linewidth]{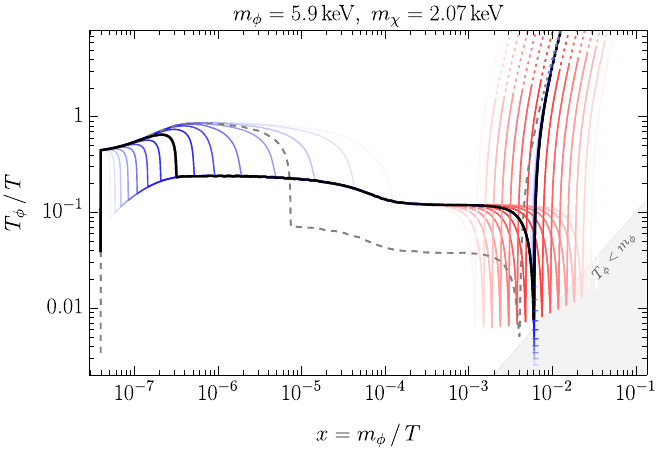}
    \hfill
    \caption{Mediator abundance (left) and its temperature ratio $T_\phi/T$ (right) plotted over inverse SM temperature $x=m_\phi/T$. Parameter variations around the denoted starting point (black curve) are shown for changes in $\lambda$ (blue) and $\kappa$ (red). Variations of $g_h$ are exemplified for a single case (gray dashed curve). Arrows in the left panel guide the eye for intuitive understanding of the impact of each parameter. Continuation beyond respective $x_\text{kd}$ is shown as dotted lines. Decay-cooling dip depths are emphasized by small horizontal dashes for variations in $\lambda$.}
    \label{fig:YandTofx}
\end{figure*}

We begin the discussion of numerical results once again with the scalar model.
The production process outlined in Sec.~\ref{sec:cooling} can be followed in Fig.~\ref{fig:YandTofx} for different choices of the couplings $\lambda$ (blue) and $\kappa$ (red) around the values $\lambda=10^{-2.78}$, $\kappa=10^{-7.24}$ (black). The left and right panels show the abundance evolution, $Y_\phi(x)$, and the ratio of dark mediator and SM temperatures, $T_\phi/T$, respectively. One additional lowered value of $g_h$ by a factor 0.1 is shown (gray dashed curve) which is instructive of the general dependence on $g_h$. All abundance curves clearly show the phases of freeze-in ($x\lesssim2\times 10^{-7}$), thermalization ($x\sim 3\times 10^{-7}$), evolution in chemical equilibrium ($x\in [3\times 10^{-7},0.006]$) and decay ($x\gtrsim0.001$). 

The QCD phase transition is visible in the temperature evolution around $x=10^{-4}$ ($T=0.1$\,GeV) but leaves $Y_\phi(x)$ unaffected. 
Different depicted values of the self-coupling $\lambda$ merely lead to earlier or later DS thermalization but do not substantially affect the produced $\phi$ abundance or late-time temperature. The decay cooling spikes in $T_\phi/T$ do deepen with larger $\lambda$ and we indicate the various overlaid minima values by a horizontal dash on each line. For very low values of $\lambda$ (not depicted), chemical equilibrium is not reached prior to decay, and the DM relic abundance is given by simply twice the freeze-in plateau, $\sim 2\times 10^{-5}$, at a correspondingly higher DM temperature.

Because decay cooling arises from the interplay of inverse cannibal reactions and decay, $\kappa$ can, to a limited degree governed by $x_\phi$, affect the final DM abundance. This is not overt from the abundance evolutions alone and would be highly unusual in, for example, superWIMP models. While superWIMP scenario may superficially seem to have similar evolution depicted in the left panel (a completed phase of mediator production, although here non-thermally, followed by a later decay), here the mediator may still be in \emph{chemical} equilibrium with itself at the time it decays, sourcing decay cooling. Stronger self-interactions support longer decay cooling and thus can increase the final DM abundance, despite the fact that the initial thermalization step remains unaltered in magnitude.  

For the chosen parameter point, decay cooling is also notably more pronounced in the temperature evolution than the thermalization step and decreases $T_\phi/T$ by more than an order of magnitude. Variations in $\kappa$ demonstrate that the decay-cooling strength depends on the available energy in the bath. It is larger for earlier decays and, as expected, stops where $T_\phi\sim m_\phi$ (gray shaded area), \emph{i.e.} once the inverse cannibal process is kinematically closed for the mean energy of dark mediators in the bath.  
Overall, this suggests that, for a given $\lambda$, earliest decays immediately after thermalization lead to stronger cooling. 

We derive above, in Sec.~\ref{sec:analytic}, that the magnitude of the thermalization step grows as $1/\sqrt{g_h}$ for smaller $g_h$ (gray dashed curve). This is the main mechanism to adjust the overall level of produced DS abundance, both, for the mediator and subsequently DM. 
Even though the cooling grows with lower $g_h$, the overall produced $\phi$-abundance still drops which actually renders the thermalization cooling contribution more relevant for \emph{heavier} DM where lower $Y_\chi$ suffice to saturate the observational energy density bound.

%----------------------------------------------------
\subsection{The SU($N$) model}
\label{sec:results-SUN}
%----------------------------------------------------

As already pointed out in Sec.~\ref{sec:model-SUN}, the non-Abelian model must be treated with care at high mediator temperatures (early and late times). We here lay out the various challenges of the non-Abelian model and discuss one benchmark evolution as a proof-of-concept of the applicability of the self-cooling mechanism.

To begin with, we consider the more constraint the model realization of gauge-multiplet DM, where mediator decay is governed by the gauge interaction, so $\Gamma_{A\to\bar\chi\chi}\propto m_A g^2 \epsilon$. This realization is under strong constraints from various sides. From numerical scans, and supported by analytic estimates, we find that  \LyA bounds, relic density constraints and sufficient life-time of $A$ corner all parts of parameter space where self-cooling contributes. \LyA bounds set a minimal $m_\chi$, which also means an upper bound on the mediator abundance via relic density constraint, 
$\Omega h^2 / (s_0 m_\chi) = Y_{\chi,0} \geq 2Y^{\max}_A \sim g_h^{3/2}$. 
Now, in order to maintain efficient $2\leftrightarrow3$ reactions, lower $g_h$ require larger $g$ which in turn shortens the lifetime of $A$ and additionally increases the expected size of thermal mass corrections. Since $m_A>2m_\chi$ is also bounded from below, one would need to fine-tune the degeneracy $\epsilon$ to many orders of magnitude to avoid $A$ decaying much faster than it can reach chemical equilibrium.  We exclude this for being highly unnatural and susceptible to quantum corrections. 
According our studies of the parameter space, this interplay leaves no room for viable self-cooling SU($2$) doublet DM in the multiplet realization.

\begin{figure}[t]
    \centering
    \includegraphics[width=0.48\textwidth]{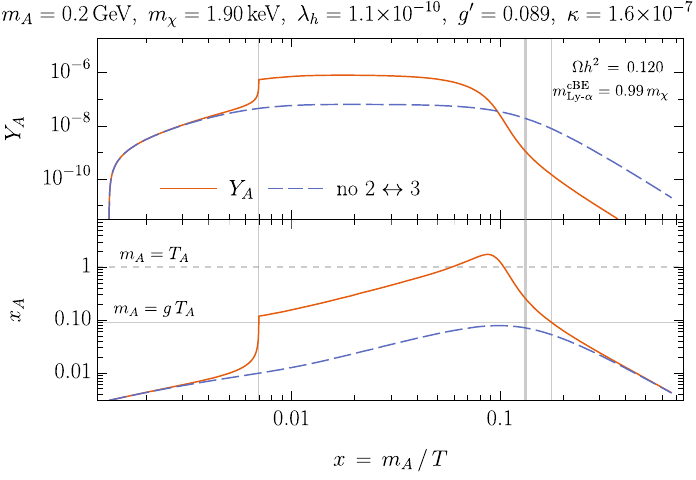}
    \caption{Evolution of abundance and inverse temperature ($Y_A$ and $x_A$) for an exemplary benchmark point in the non-Abelian model (orange lines) together with corresponding result for the same parameter point with disabled $2\leftrightarrow3$ scatterings (blue dashed). \LyA bounds obtained from the cBE evolution until $x_{f,\text{cBE}}=0.13$ (thick gray vertical line) give $\mLyA^\text{cBE}=0.99m_\chi$.}
    \label{fig:SUNevolution}
\end{figure}

Fig.~\ref{fig:SUNevolution} shows the evolution of $Y_A$ and $x_A$ (solid and dashed curves) for a benchmark DM model for the independent non-Abelian realization where $\kappa\ll g$. The horizontal gray line indicates $x_A=g$, which means that here the Debye-mass term $\Delta m_A(T_A)\sim m_A$ and above it thermal mass corrections begin to be subdominant. Since chemical equilibrium is maintained until $\Delta m_A(T_A)\ll m_A$, we do not expect large impact from thermal corrections to this benchmark point. The reason is that any kinetic energy gained from decreasing Debye mass terms will immediately thermalize, eventually leading to the same thermal distribution as if $\Delta m_A$ was neglected from the start. Since $\Delta m_A(T_A)\sim g T_A \lesssim g m_h \ll m_h$, one expects no phase-space suppression from degeneracy during freeze-in unless $m_A\approx m_h/2$ itself. 

Overall, the non-Abelian model phenomenology observed in Fig.~\ref{fig:SUNevolution} is largely identical to the scalar model, \emph{cf.}~Fig.~\ref{fig:YandTofx}.
Both, the thermalization step and decay cooling, are distinctly visible and follow identical physics as was explained in Sec.~\ref{sec:results-evolution}. 
Viable parameter space with sizable and distinct self-cooling features is found for rather small gauge couplings, $g\sim 0.1-0.01$, with larger couplings mainly limited by thermal corrections, perturbativity in the theoretical calculation and stability of the numerical solutions.

Decay cooling is present but subdominant in the given benchmark point. Since $m_A$ now also governs the times when $\Delta m_A (T_A)\sim m_A$ (thin gray vertical lines), it can no longer be altered  without concern as easily as in the scalar model. Moreover, $x_A\gtrsim1$ is reached in Fig.~\ref{fig:SUNevolution}, thus variations of $m_A$ around the presented parameter point cannot be absorbed into changes of other parameters as we mentioned above to be the case for fully-relativistic evolutions.

The late-time treatment of and beyond kinetic decoupling at rising temperatures, as discussed in Sec.~\ref{sec:KD}, leads to challenges for the non-Ablelian model when estimating $x_\text{kd}$ arising from the momentum dependence of the cubic gauge-boson vertex for the massive vector bosons in our effective theory. The naive elastic scattering rate, $\Gamma_\text{el} \approx n_A\times \left< \sigma v   \right>^\text{transfer}_{2\to2} $ is still much larger than $H$ when decay cooling stops. Hence, this naive estimate predicts kinetic decoupling many orders of magnitude later, which means the mean effective velocity will be dominated by tiny, hot amounts of the DM abundance frozen in at latest times. We see several challenges in this setup:  
    Firstly, the naive elastic scattering rate estimate may be inaccurate, if the momentum exchange per scattering is small. Rather, the momentum-reshuffling among bath particles needs to be large per Hubble-time.
    Secondly, if the DS heats too much, our effective description will break down as UV effects such as higher-dimension effective vertices, become important. Alternatively, unitarization of $2\to2$ scatterings becomes of concern, see Sec.~\ref{sec:model-SUN}. 
    Thirdly, $A$ gains a thermal mass $\Delta m_A = \sqrt{N/3}\, g \, T_A \lesssim g\, m_h$, which we presently do not account for. Since energy is conserved within the DS, we do not expect early-time thermal mass corrections to be significant, as long as there is a subsequent period of time where chemical equilibrium is maintained, $A$ does not yet decay and $m_A \gg \Delta m_A$. 
    Lastly, even if all above conditions lead to very late kinetic decoupling, eventually the statistical cosmological evolution should not be defined by few high-energetic modes and WDM simulations do no longer provide a reasonable reference value.

Given these challenges regarding the late-time treatment in the non-Abelian model, our analysis stopped short of a complete MCMC scan. 
The benchmark model of Fig.~\ref{fig:SUNevolution} yields a viable DM candidate with $m_\chi=1.9$\,keV, similar  to the scalar case. By choosing $x_\text{kd}=0.13$ (thick vertical gray line) and using free-streaming results beyond that, we find  $m_\chi>\mLyA=1.88$\,keV. We emphasize that this choice of $x_\text{kd}$ is purely heuristic, informed from expectations built in the scalar model, and may be subject to sizable modifications from any or all of the challenges.

We remark here, that there are clear paths forwards for improvement: A more reliable estimate of the time of kinetic decoupling may be accessible from detailed study of the momentum change rate. A dedicated partial wave analysis to investigate unitarity bounds is well defined, while a concrete choice for a UV-completion allows to match higher-order operators. Also the treatment of dynamic Debye-masses in the cBE is accessible.

These improvements are beyond the scope of the present, explorative work and we leave a robust study of kinetic decoupling in the non-Abelian model for future investigations.

%----------------------------------------------------
\subsection{MCMC parameter scan}
\label{sec:MCMC}
%----------------------------------------------------

\begin{figure}[t]
\centering
\includegraphics[width=0.42\textwidth]{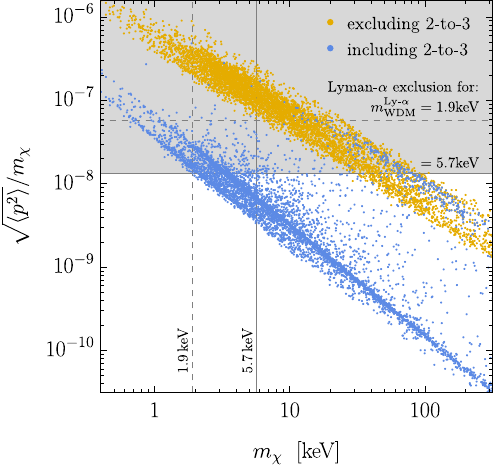}
\caption{The results from the MCMC scan satisfying $\Omega h^2=0.12\,\pm\,3\%$ (blue points) together with corresponding points (yellow) without self-interactions  but rescaled portal coupling $g_h$ in order to give the same relic abundance. The gray shaded region is excluded by the \LyA data recast from the $m_\text{WDM}^{\text{Ly-}\alpha}=5.7$\,keV limit ($1.9$\,keV boundary shown dashed). 
The bottom left quadrant thus includes points which are allowed but lighter than the WDM bound.
}
\label{fig:MCMC}
\end{figure}

We return to the scalar model for the remainder of this chapter. To explore the whole parameter space, we employ MCMCs to identify parameter regions of interest. 
We target parameter points which saturate the DM energy density, $\Omega h^2=0.12\,\pm\,10\%$, and lie around the exclusion limit for \LyA bounds ($\pm\,10\%$). To exert better control over which parameter ranges are being searched, we refrain from altering $m_\chi$ to match the correct DM energy density, as $m_\chi$ does not affect the DS evolution except for degenerate models, and adjust only $g_h$.

We scan over the parameter space, with  masses and couplings
\begin{gather}
    m_\phi \in [10^{-3},10^{3}] {\,\rm {MeV}}, ~~ m_\chi\in [10^{-3},10^{2}] {\,\rm {MeV}}, \nonumber\\
    g_{h}\in [10^{-11},10^{-6}],  ~\lambda\in [10^{-4},10^{-1}], ~\kappa\in[10^{-10},10^{-4}], \nonumber
\end{gather}
with logarithmic sampling and condition $m_\phi>2m_\chi$.
A subset of 6327 points, satisfying $\Omega h^2=0.12$ within $\pm3\%$, is shown in Fig.~\ref{fig:MCMC}, plotting normalized velocity dispersion $\sqrt{\langle p^2\rangle}/m_\chi$ on the vertical axis, which can be directly compared to re-cast \LyA limits, \emph{cf.}~Eq.~\eqref{eq:LyA}. The entire gray region is excluded when adopting $m_\text{WDM}^{\text{Ly-}\alpha}=5.7$\,keV. A dashed horizontal line indicates how this limit would change for $m_\text{WDM}^{\text{Ly-}\alpha}=1.9$\,keV instead. Blue points are direct result of the scan, while yellow ones are their counterparts with cannibal reactions neglected and rescaled portal coupling $g_h$ to compensate for the change in relic abundance.

The plot demonstrates how the inclusion of number-changing self-interactions shifts the experimental bounds from \LyA observations, even by multiple orders of magnitude.
Additionally, comparing with the dashed vertical lines indicating the masses of 5.7 and 1.9\,keV, it is evident that many of the points that would have been excluded by \LyA observations, are allowed due to self-cooling. Moreover, there are a number of points found in the scan with masses significantly lower than allowed in the WDM model. In particular, with the default $m_\text{WDM}^{\text{Ly-}\alpha}=5.7$\,keV limit, the lowest DM mass found by our MCMC searches which saturates the DM energy density and still remains in agreement with \LyA bounds has $m_\chi=1.92$\,keV. Therefore, it was found that the self-cooling mechanism for the scalar model allows to undercut the often quoted WDM bound on $m_\chi$ by roughly a factor 3. In fact, the self-cooling is notably stronger than that and often decreases the obtained velocity dispersion by factors of $\sim 200$, as is apparent from the vertical separation of the blue and yellow point ensembles. This is because \LyA bounds for freeze-in DM are typically more stringent when compared to WDM, for minimal models around $16$\,keV \cite{Decant:2021mhj}, and thus almost a factor 10 larger than the lightest viable parameter point we found.

%----------------------------------------------------
\subsection{Parameter space analysis}
\label{sec:results-2d}
%----------------------------------------------------

\begin{figure}[t]
\centering
\includegraphics[width=0.99\linewidth]{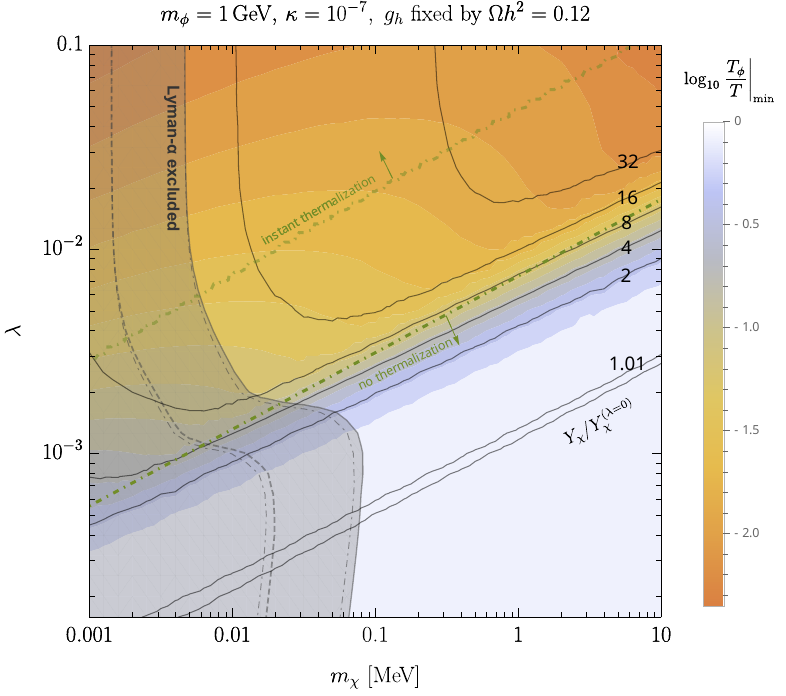}\\[2pt]
\includegraphics[width=0.99\linewidth]{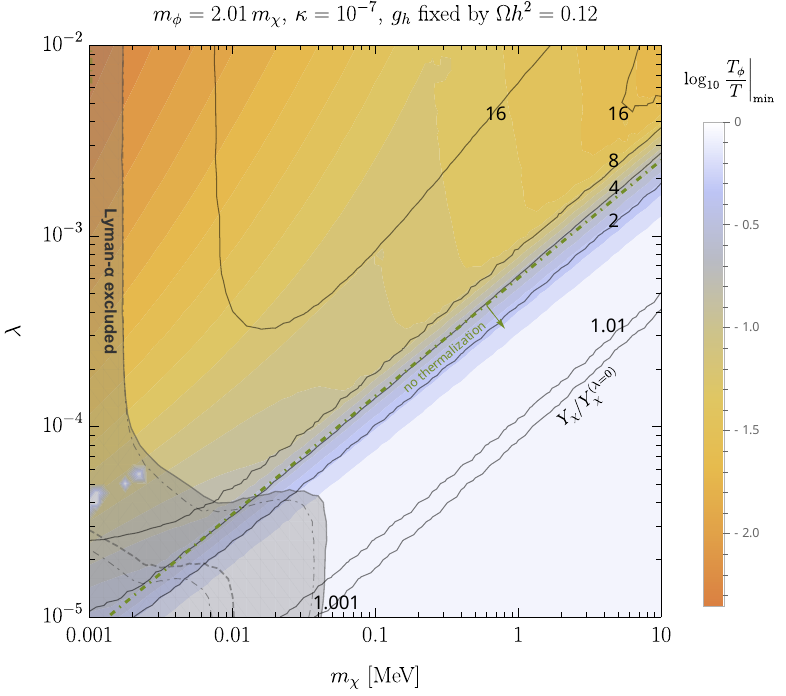}
\caption{The \LyA bounds, gray shaded area, superimposed on a colored contour plot showing the maximal size of the self-cooling effect throughout the evolution. The black solid contours indicate the relative boost in relic density due to $2\leftrightarrow3$ scatterings. The gray dashed curve shows how the \LyA exclusion boundary for $m_\text{WDM}^{\text{Ly-}\alpha}=1.9$\,keV instead of $5.7$\,keV. (For each choice dash-dotted lines neglect contributions beyond kinetic decoupling.) 
The top and bottom panels show constant $m_\phi=1$\,GeV and near-degenerate $m_\phi=2.01m_\chi$, respectively.
}
\label{fig:2D}
\end{figure}

To provide better understanding of parametric dependence of the scalar model, we present a scan of the parameter space in $m_\chi$ and $\lambda$, while adjusting $g_h$ to saturate the observed DM energy density, in Fig.~\ref{fig:2D}.  

We show results for fixed $m_\phi=1$\,GeV (top panel) and the degenerate definition $m_\phi\equiv 2.01\times m_\chi$ (bottom panel). The regions excluded by \LyA bounds are covered by a gray area and the dashed line again indicates change to that exclusion if $m_\text{WDM}^{\text{Ly-}\alpha}=1.9$\,keV were adopted. For each choice of $m_\text{WDM}^{\text{Ly-}\alpha}$ we also show a dash-dotted line which indicates the result when neglecting contributions beyond kinetic equilibrium, \emph{cf.}~Sec.~\ref{sec:KD}. At the bottom of both plots, $\lambda$ is small enough that $2 \rightarrow 3$ processes are never efficient. On the other hand, at $\lambda\sim 10^{-2}-10^{-1}$ they ensure essentially instant internal DS thermalization already during freeze-in from SM. Both of these limits are indicated on Fig.~\ref{fig:2D} as green lines, in between which the cEQ is reached at some intermediate time in the $\phi$'s evolution (Fig.~\ref{fig:EvolutionOverview} shows an example of this). 
As expected, rapid changes in DS cooling relative to SM and also DM abundance increase over the non-inverse-cannibalizing result are found at the boundary where no thermalization is reached.

The shape of the exclusion contour is explained as follows. At lowest $\lambda$, no $2 \rightarrow 3$ processes are active. As $\lambda$ increases it becomes large enough for the mediator to cool from thermalization, at which point \LyA bounds abruptly weaken allowing lower values of $m_\chi$. As $\lambda$ increases further, the secondary decay cooling phase lowers \LyA further. At largest $\lambda$, the mediator cools all the way to non-relativistic temperatures, $x_\phi\gtrsim1$, at which point the $2\to3$ closes kinematically (rather than being suppressed by dilute $Y_\phi$) and \LyA bounds are no longer affected by changes in $\lambda$.
Overall, it is worth stressing that for both cases the self-cooling effect alleviates the bound on $m_\chi$ by more than an order of magnitude.

In addition to the \LyA exclusion limits, both panels show shaded contours of the maximal cooling strength, $\min\{T_\phi/T\}$ (color coded from blue - lowest, to orange - largest cooling effect) as well as the increase in total produced DM abundance due to the inclusion of $\Gamma_{2\to3}$ in the solution of the cBE, labeled $Y_\chi/Y_\chi^{(\lambda\to0)}$ (black solid contours). The mediator temperature changes rapidly within the region where the mediator first thermalizes. In both panels, the bottom right corner never thermalizes, while decay-cooling contributes towards the top-left corner of the parameter plane where self-interactions are strong and low $m_\chi$ dictate larger $g_h$ and, thus, smaller thermalization effects.  Towards the top-right corner in the degenerate model scenario (bottom panel), $x_\phi\gtrsim 1$ is reached prior to decay and $Y/Y^{(\lambda\to0)}$ defines a hill in the center of the plot, on top of which thermalization-cooling dominates.

Still, $\mLyA/\tilde{m}_{\text{Ly}\alpha}$ increases continuously towards the top-left corner because $m_\phi\propto m_\chi$. 
Once the black contours of $Y_\chi/Y_\chi^{(\lambda=0)}$ become independent of $\lambda$, \emph{e.g.}~$Y_\chi/Y_\chi^{(\lambda\to0)}=16$ near $m_\chi\sim 0.01$\,MeV, there is no more cooling gained from  thermalization at larger $\lambda$ and additional reduction of $T_\phi/T$ found along a fixed contour is driven by decay cooling.
We thus find that $\min\{T_\phi/T\}$ by itself does not serve as a useful indicator for the loosening of \LyA bounds as it cannot account for decay cooling. 

As a detail, the exclusion contour in the fixed mass scenario (upper panel) is not independent of very small $\lambda<10^{-3}$, showing $\mathcal{O}(5\%)$ dependence from $10^{-4}$ to $10^{-3}$. We attribute this to minor self-cooling arising near the time of decay. The fact that \LyA bounds grow stronger as $\lambda$ slowly increases at low values appears to be contradictory to the discussion in previous chapters. The difference is caused by an impact of non-relativistic physics. As $\lambda$ increases, $g_h$ drops because $Y_\phi\propto Y_\chi$ must remain constant on the contour, thus the onset of self-cooling remains near $x_\text{dec}$ over a wider window in $\lambda$. For the given $m_\phi=1$\,GeV, the mediator is already close to the non-relativistic regime. Even at small $\lambda$, $\max\{x_\phi\}\sim 0.1$. Additional cooling causes the model to evolve closer to a fully non-relativistic phase for a longer time and the weaker redshift of $\rho_\phi^\text{NR}$ outweighs the cooling.  
Only once $\lambda$ becomes large enough to source thermalization at much earlier times  is $T_\phi$ sufficient to fully thermalize $\phi$.
We note that our precision in $\Omega h^2$ is at the level of $0.5\%$ but fluctuations thereof appear uncorrelated to the observed contour shape.

%%%%%%%%%%%%%%%%%%%%%%%%%%%%%%%%%%%%%%%%%%%%%%%%%%%%%
\section{Conclusions}
\label{sec:conclusions}
%%%%%%%%%%%%%%%%%%%%%%%%%%%%%%%%%%%%%%%%%%%%%%%%%%%%%

In this work we uncovered a new mechanism for cooling down relic populations that can significantly alleviate the \LyA bounds of a light dark matter candidate.  The dynamics of the complete self-cooling process is based on inverse cannibal $2 \rightarrow 3$ reactions of a generic mediator particle within the freeze-in production framework.

We provide an analytic understanding of the underlying physics of the mechanism, identifying  two phases of self-cooling: thermalization within the dark sector and decay cooling. The former is due to the $2 \rightarrow 3$ processes exchanging kinetic energy for mass while driving the system towards chemical equilibrium. Hence, this process can serve as cooling if the mediator particle starts out at above-thermal temperatures, given its number density. Decay cooling occurs from simultaneously efficient inverse cannibal- and decay-processes and, interestingly, has been found to have a stronger impact on \LyA bounds than self-thermalization via cannibal-reactions itself in large regions of parameter space.

We exemplified self-cooling on two concrete models and studied its effectiveness. \LyA bounds can be modified by several orders of magnitude in the sub-MeV region and, pushing for smallest masses, frozen-in dark matter saturating the observed energy density can be found as light as $1.9 (0.7)\text{\,keV}$, without being in conflict with \LyA observations. Self-cooling substantially alleviates the bounds compared to the $ 5.7 (1.9)$\,\text{keV} warm dark matter limits determined from simulations and $ 16 (4.1)\text{\,keV}$ for a minimal freeze-in realizations. Within the exact context of the studied models self-cooling can be as effective as lowering the mass limit by more than two orders of magnitude. Our results corroborate that colloquially known WDM bounds are not to be applied indiscriminately to differently produced DM models.

In calculating the \LyA bounds, we identified and explained a spurious enhancement of the mean dark matter velocity dispersion. It arises when second-moment Boltzmann equations, which inherently assume kinetic equilibrium, are carelessly extended to much later times. This is a feature of the temperature evolution and barely noticeable in produced relic densities. A dedicated treatment found that decays after kinetic decoupling contribute only a minor correction to the velocity dispersion in the investigated model.

Our results corroborate the difference between higher-order and leading-order processes in QFT. Even though a process may be small for being ``higher-order" in traditional power counting of perturbative QFT, if it opens a novel reaction process, it may still have to be included to obtain reliable results. Especially so, for processes that break otherwise conserved thermodynamic quantities. In our case $2\to3$ reactions break number conservation and allow for internal thermalization, hence, simplistic power-suppression arguments must not serve as a basis for neglecting collision terms in a Boltzmann equation. Instead only processes which are justifiably negligible or irrelevant in the context of early Universe thermodynamics may be disregarded. For our simple model setup, the leading-order $2\to3$ scattering rate scales as the quartic coupling cubed, $\lambda^3$, yet still sources strong self-cooling starting from small $\lambda\sim10^{-4}$.

Beyond our present work, the cooling mechanism is sufficiently general that it may well be possible to incorporate it into more complicated, phenomenologically motivated models  with relative ease.
From this point of view, broken non-Abelian gauge theories are an attractive candidate for number-changing interactions, as they necessarily arise from the three-point gauge vertex. However, we find a proper treatment of kinetic decoupling in such theories to host several hurdles related to UV physics, unitarity, thermal masses and repeated small-angle scatterings. Most of these challenges are not novel by themselves and there are already clear paths forwards to a better understanding of self-cooling in non-Abelian theories. 

As current model building efforts become more intricate, most models involve extended, multi-field dark sectors. Here, we see fertile ground for further applications of the presented cooling mechanism.

%%%%%%%%%%%%%%%%%%%%%%%%%%%%%%%%%%%%%%%%%%%%%%%%%%%%%
\subsection*{Acknowledgments}
We thank Mathias Garny for helpful discussions in early stages of this work. 
This work was supported by the National Science Centre (Poland) under the research Grant No. 2021/42/E/ST2/00009.

%%%%%%%%%%%%%%%%%%%%%%%%%%%%%%%%%%%%%%%%%%%%%%%%%%%%%
\bibliography{biblio}{}

\end{document}